\RequirePackage{lineno} 
\documentclass[a4paper,10pt]{article}

\usepackage{amsmath}
\usepackage{amsthm}
\usepackage{amsfonts}
\usepackage{hyperref}
\usepackage{color}
\usepackage{multirow}
\usepackage[dvips]{graphicx}
\usepackage{subfigure}
\usepackage{amssymb}
\usepackage{epsfig}

\usepackage{setspace}

\newcommand{\comment}[1]{}
\numberwithin{equation}{section}

\begin{document}



\bibliographystyle{plain}

\title{\textbf{An Alternative Method to Implement Contact Angle Boundary Condition on Immersed Surfaces for Phase-Field Simulations}}

\author{$\textbf{Jun-Jie Huang}^{1, 2,  3\footnote{Corresponding author. E-mail: jjhuang1980@gmail.com; jjhuang@cqu.edu.cn.}}, 
\textbf{Jie Wu}^{4}$\\
\\
$ ^1$  Department of Engineering Mechanics, College of Aerospace Engineering, \\
Chongqing University, Chongqing 400044, China \\
$ ^2$ Chongqing Key Laboratory of Heterogeneous Material Mechanics \\
(Chongqing University), Chongqing 400044, China\\
$ ^3$ State Key Laboratory of Mechanical Transmission, \\
Chongqing University, Chongqing 400044, China \\
$ ^4$ Department of Aerodynamics, Nanjing University of Aeronautics and Astronautics, \\
Yudao Street 29, Nanjing, Jiangsu 210016, China}

\maketitle

\textbf{Abstract}

In this paper, we propose an alternative approach to implement the contact angle boundary condition on immersed surfaces
for phase-field simulations of two-phase flows using the Cahn-Hilliard equation on a Cartesian mesh.
This simple and effective method was inspired by previous works on the geometric formulation of the wetting boundary condition.
In two dimensions, by making full use of the hyperbolic tangent profile of the order parameter, 
we were able to obtain its unknown value at a ghost point from the information at \emph{only one} point in the fluid. 
This is in contrast with previous approaches using interpolations involving several points.
The special feature allows this method to be easily implemented on immersed surfaces (including curved ones)
that cut through the grid lines. It is verified through the study of two examples: 
(1) the shape of a drop on a circular cylinder with different contact angles;
(2) the spreading of a drop on an embedded inclined wall with a given contact angle.

\textbf{Keywords}: 
\textit{Wetting Boundary Condition},
\textit{Immersed Surface},
\textit{Phase-Field Simulation},
\textit{Cahn-Hilliard Equation}.


\section{Introduction}\label{sec:intro}

Two-phase flows with moving contact lines (MCLs) are commonly encountered
in various areas of our daily life and a number of industries~\cite{deGennes2004}.
Numerical simulations of such flows have had significant development recently~\cite{arfm14-mcl}.
One of the widely used methods is the phase-field (or diffuse-interface) method~\cite{DIMReview98},
which uses a physically defined indicator function, known as the order parameter (or the concentration),
to differentiate the two fluids.
Its evolution is usually described by the Cahn-Hilliard equation (CHE)~\cite{jacqmin99jcp, jcp03pfm}.
An important component in phase-field simulations is the boundary condition 
on a solid surface with specified wettability characterized by a contact angle.
There are several formulations for the wetting boundary condition (WBC) 
and some comparative studies of them may be found in~\cite{ijnmf15-wbc-pfs}.
Among various formulations of the WBC, 
the geometric formulation possesses the property to always keep the local contact angle
to be the specified value~\cite{pre07-geom-wbc, cf11-accur-cabc}.
As far as we know, most of previous studies using the geometric formulation on a Cartesian mesh
implemented the WBC on a flat surface
that form one of the boundaries of the simulation domain.
The way to implement it on a general solid surface (including curved ones)
immersed inside the domain has been rarely investigated.
An exception is the very recent work~\cite{jcp15diim-curved} 
which extends the characteristic interpolation (CI) originally proposed in~\cite{cf11-accur-cabc} to curved surfaces.
However, when the original CI is employed,
one has to determine the local configuration of the interface
before applying any interpolation formula to obtain the unknown value (of the order parameter) along the characteristic line
(see~\cite{jcp15diim-curved} for details).
This is mainly because at different places
an immersed curved surface has different orientations and intersects the Cartesian mesh in different ways.
The main purpose of this work is to propose an alternative approach to implement the WBC 
on general immersed surfaces that is able to avoid such complex issues.
The key idea is to make full use of the hyperbolic tangent profile of the order parameter
(as will be illustrated later).
The remainder of this work is organized as follows.
Section \ref{sec:method} introduces the theoretical and numerical methodology
with the emphasis on the newly proposed method.
Section \ref{sec:res-dis} presents the studies of two examples to verify the new approach.
Section \ref{sec:conclusion} concludes this paper.

\section{Theoretical and Numerical Methodology}\label{sec:method} 


The present simulation of two immiscible fluids uses
a hybrid lattice-Boltzmann finite-difference approach~\cite{hybrid-mrt-lb-fd-axisym}.
The Cahn-Hilliard equation (CHE) for the interface dynamics
is solved by a finite-difference method 
(using $2^{nd}-$order isotropic schemes for spatial derivative evaluation)
with the $4^{th}-$order Runge-Kutta time marching scheme.
The hydrodynamics of the flow is simulated by the lattice-Boltzmann method (LBM)
(the current version is mainly based on~\cite{cma09lee-para-curr, jcp10lbm-drop-impact}).
Most specifics of the present method may be found in~\cite{hybrid-mrt-lb-fd-axisym, cma09lee-para-curr, jcp10lbm-drop-impact}.
For completeness, some key components are briefly recaptured below first.
The newly proposed WBC will then be introduced in detail.


For a system of binary fluids, 
a free energy functional $\mathcal{F}$ may be defined as, 
\begin{equation}
  \label{eq:fe-functional-def}
  \mathcal{F} (\phi, \boldsymbol{\nabla} \phi)
  = \int_{V} \bigg( \Psi (\phi) 
  + \frac{1}{2} \kappa \vert \boldsymbol{\nabla} \phi \vert ^2 
  \bigg) dV ,
\end{equation}
where $\Psi (\phi)$ is the bulk free energy density, 
and the second term 
is the interfacial energy density ($\kappa$ is a constant). 
With $\Psi (\phi)$ chosen to be in the double-well form, 
$\Psi (\phi) = a (\phi^{2} - 1)^{2}$ ($a$ is a constant),
the order parameter $\phi$ varies between $1$ in one of the fluids 
(named \textit{fluid A} for convenience) and $-1$ in the other (named \textit{fluid B}).
The chemical potential $\mu$ is obtained
by taking the variation of $\mathcal{F}$ with respect to $\phi$,
\begin{equation}
  \label{eq:chem-potential}
\mu = \frac{\delta \mathcal{F}}{\delta \phi} 
= \frac{d \Psi (\phi)}{d \phi} -  \kappa \nabla ^2 \phi
= 4 a \phi (\phi ^2 -1) -  \kappa \nabla ^2 \phi  . 
\end{equation}
The coefficients $a$ and $\kappa$ are related to the interfacial tension $\sigma$ and 
interface width $W$ as
$a = \frac{3 \sigma}{4 W}$ and $\kappa = \frac{3 \sigma W}{8}$
~\cite{ijnmf09-pflbm-mobility}.
Across a flat interface in equilibrium $\phi$ is described by a hyperbolic tangent profile 
$\phi (\zeta) = - \tanh \frac{\zeta - \zeta_{0}}{W / 2}$
where $\zeta$ is the coordinate along the axis normal to the interface,
and $\zeta_{0}$ is the position where $\phi = 0$.
With contribution due to convection taken into account, the CHE reads~\cite{jacqmin99jcp},
\begin{equation}
  \label{eq:che}
  \frac{\partial \phi}{\partial t}
  + \boldsymbol{u} \cdot \boldsymbol{\nabla} \phi
  = \boldsymbol{\nabla} \cdot (M  \boldsymbol{\nabla} \mu)   ,
\end{equation}
where $M$ is the (constant) mobility, and $\boldsymbol{u}$ is the local fluid velocity.

When the single-relaxation-time collision model is used,
the lattice-Boltzmann equations (LBEs) read~\cite{cma09lee-para-curr},
\begin{equation}
\label{eq:lbe-vardv}
f_{i} (\boldsymbol{x} + \boldsymbol{e}_{i} \delta _t, t + \delta _t ) - f_{i} (\boldsymbol{x} , t)
= - \frac{1}{\tau_{f}} (f_{i} - f_{i}^{eq})
+ \bigg( 1 - \frac{1}{2 \tau_{f}} \bigg) (\boldsymbol{e}_{i} - \boldsymbol{u} ) \cdot 
[ \boldsymbol{\nabla} \rho c_{s}^{2} (\Gamma_{i} - \Gamma_{i} (0)) - \phi \boldsymbol{\nabla} \mu \Gamma_{i} ] ,
\end{equation}
where $f_{i}$ and $f_{i}^{eq}$ are the distribution functions (DFs) and equilibrium DFs
along the direction of the lattice velocity $\boldsymbol{e}_{i}$ ($i = 0, 1, \cdots, b$),
$w_{i}$ is the weight for the direction along $\boldsymbol{e}_{i}$,
$c_{s}$ is the lattice sound speed,
$\tau_{f}$ is the relaxation parameter 
which is related to the dynamic viscosity as $\eta = \rho c_{s}^{2} (\tau_{f} - 0.5) \delta_{t}$.
The equilibrium DFs are given by,
\begin{equation}
f_{i}^{eq} = w_i \bigg[ p + \rho c_s^2 \bigg( \frac{1}{c_s^2}  e_{i \alpha} u_{\alpha}
+ \frac{1}{2 c_s^4} (e_{i \alpha} e_{i \beta} - c_s^2 \delta_{\alpha \beta}) u_{\alpha} u_{\beta} \bigg) \bigg] ,
\end{equation}
where $p$ is the hydrodynamic pressure,
and the term $\Gamma_{i}$ 
is given as
$\Gamma_{i} (\boldsymbol{u}) = w_{i} [ 1 +  \frac{1}{c_s^2}  e_{i \alpha} u_{\alpha}
+ \frac{1}{2 c_s^4} (e_{i \alpha} e_{i \beta} - c_s^2 \delta_{\alpha \beta}) u_{\alpha} u_{\beta} ]$.
The common D2Q9 lattice velocity model ($b=8$) is adopted for the current two dimensional study 
(details of D2Q9 may be found in~\cite{pre2000-lbe-theory}).
The macroscopic variables are calculated as,
\begin{equation}
p =  \sum_{i} f_{i} +  \frac{1}{2} \delta_{t}  (\boldsymbol{u} \cdot  \boldsymbol{\nabla} \rho c_{s}^{2}),
\end{equation}
\begin{equation}
\rho \boldsymbol{u} = \frac{1}{c_{s}^{2}} \sum_{i} f_{i} \boldsymbol{e}_{i} -  \frac{1}{2} \delta_{t} ( \phi \boldsymbol{\nabla} \mu) .
\end{equation}
Through the Chapman-Enskog analysis, 
it can be found that the LBEs, Eq. (\ref{eq:lbe-vardv}),
approximate the following equations at the macroscopic level~\cite{jcp10lbm-drop-impact},
\begin{equation}
\label{eq:cnsecontinuity-vardv}
\frac{ \partial p}{ \partial t} + \rho c_{s}^{2} \boldsymbol{\nabla} \cdot \boldsymbol{u} = 0 ,
\end{equation}
\begin{equation}
\label{eq:cnsemomentum-vardv}
\rho \bigg( \frac{ \partial \boldsymbol{u}}{ \partial t} 
+  \boldsymbol{u} \cdot \boldsymbol{\nabla} \boldsymbol{u} \bigg)
= - \boldsymbol{\nabla} p - \phi \boldsymbol{\nabla} \mu + \boldsymbol{\nabla} \cdot  \mathbf{\Pi} ,
\end{equation}
where $\mathbf{\Pi}$ is the common viscous stress tensor for Newtonian fluids.
To improve stability, we actually used the multiple-relaxation-time collision model~\cite{pre2000-lbe-theory}.
On an immersed solid wall, the bounce-back condition is applied for the DFs $f_{i}$
according to the simple and efficient method given in~\cite{pof01lbm-bc}.


Eqs. (\ref{eq:chem-potential}) and (\ref{eq:che})
require suitable boundary conditions.
For the chemical potential $\mu$, we employ the no-flux condition on a surface $S$,
\begin{equation}
\label{eq:mu-bc}
\boldsymbol{n}_{S} \cdot \boldsymbol{\nabla} \mu \vert _{S} = \frac{\partial \mu}{\partial n_{S}} \bigg \vert _{S}  = 0 ,
\end{equation}
where $\boldsymbol{n}_{S}$ denotes the unit normal vector on $S$ pointing \textit{into} the fluid.
The boundary condition for $\phi$ basically follows the geometric formulation~\cite{pre07-geom-wbc, cf11-accur-cabc}.
It is assumed that the contours of $\phi$ in the diffuse interface are parallel to each other everywhere.
Besides, we also assume that in the direction normal to the interface, 
the hyperbolic tangent profile of $\phi$ is well preserved (or just slightly perturbed).
The direct use of this profile makes the current method differ from previous works
(e.g., ~\cite{pre07-geom-wbc, cf11-accur-cabc, jcp15diim-curved}).

\begin{figure}[htp]
  \centering
 \includegraphics[trim = 0mm 0mm 0mm 0mm, clip, scale = 1.0, angle = 0]{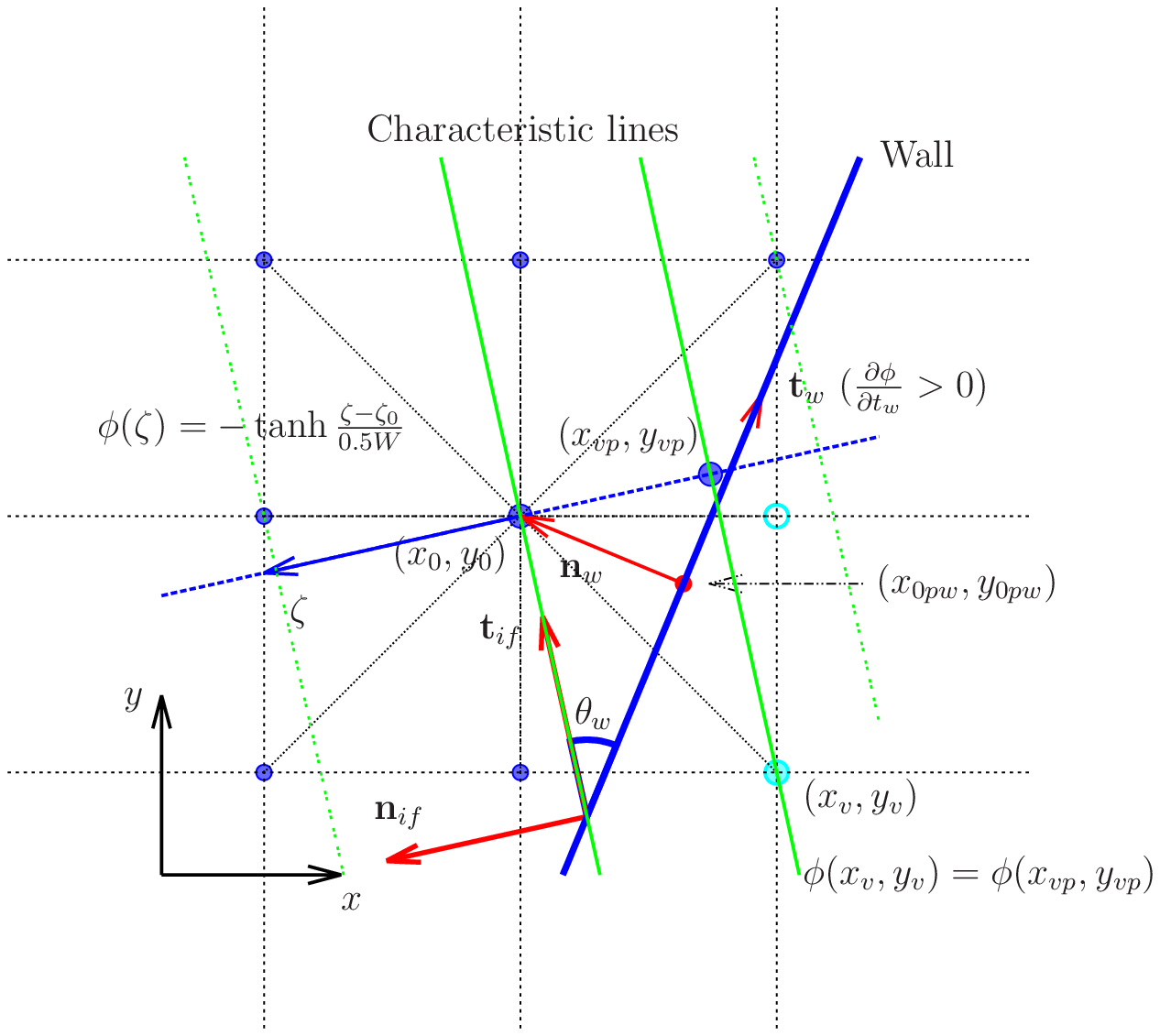}
  \caption{Illustration of the proposed implementation of the wetting boundary condition for the order parameter $\phi$
  on an immersed wall with a contact angle $\theta_{w}$.}
  \label{fig:projection-nif}
\end{figure}

Figure \ref{fig:projection-nif} illustrates
the implementation of the boundary condition for $\phi$ on an arbitrarily positioned wall 
with a specified contact angle $\theta_{w}$ immersed in the fluid domain.
We consider a point $(x_{0}, y_{0})$ in the fluid near the wall.
When the grid is fine enough to resolve the largest curvature of the wall,
it is reasonable to assume that locally the wall can be approximated by a straight line.
The eight points surrounding $(x_{0}, y_{0})$ 
may be divided into two groups:
those on the same side of the line as $(x_{0}, y_{0})$ (Group-S) and those on the other side of the line (Group-O).
The WBC is used to obtain the values of $\phi$ at the ghost points belonging to Group-O, 
which are required to calculate the derivatives
$\frac{\partial \phi}{\partial x}$, $\frac{\partial \phi}{\partial y}$ and $\nabla^{2} \phi$ 
at $(x_{0}, y_{0})$.
We take one of the ghost points, $(x_{v}, y_{v})$ in fig. \ref{fig:projection-nif}, as an example.
The specific steps to obtain $\phi (x_{v}, y_{v})$ are described below.
For convenience, a vector $\boldsymbol{n} = (\cos \theta, \sin \theta)$ in the $x-y$ plane 
is sometimes represented by a complex variable $e^{i\theta} = \cos \theta + i \sin \theta$.
\begin{itemize}
\item{Step 1: To determine the local directions that are normal and tangential to the wall, $\boldsymbol{n}_{w}$ 
and $\boldsymbol{t}_{w} = \boldsymbol{n}_{w} e^{ i (-\frac{\pi}{2}) }$,
where $\boldsymbol{n}_{w} = (x_{0} - x_{0pw}, y_{0} - y_{0pw}) / s_{0}$,
the point $(x_{0pw}, y_{0pw})$ is the projection of $(x_{0}, y_{0})$ onto the wall, and $s_{0}$ is the distance between
$(x_{0pw}, y_{0pw})$ and $(x_{0}, y_{0})$.}
\item{Step 2: To calculate the local tangential gradient of $\phi$, $\frac{\partial \phi}{\partial t_{w}}$, 
using the values of $\phi$ at (some of) the points in Group-S, 
so as to determine the direction of the characteristic lines parallel to the interface: 
$\boldsymbol{t}_{if} = \boldsymbol{t}_{w} e^{i \theta_{w}}$ if $\frac{\partial \phi}{\partial t_{w}} > 0$ (as in fig. \ref{fig:projection-nif})
and $\boldsymbol{t}_{if} = \boldsymbol{t}_{w} e^{i (\pi - \theta_{w})}$ if $\frac{\partial \phi}{\partial t_{w}} < 0$
(another situation not shown here).}
\item{Step 3: To calculate the direction normal to the characteristic lines 
$\boldsymbol{n}_{if} = \boldsymbol{t}_{if} e^{i \frac{\pi}{2}}$ 
(note the gradient of $\phi$ along $\boldsymbol{n}_{if}$ is ensured to be negative).}
\item{Step 4: To project the point $(x_{v}, y_{v})$ to the line passing through $(x_{0}, y_{0})$ 
along the direction $\boldsymbol{n}_{if}$
to obtain the point $(x_{vp}, y_{vp})$, and to further calculate its coordinate on the $\zeta-$axis in fig. \ref{fig:projection-nif} 
(denoted as $\zeta_{vp}$).
Note that the projection of $(x_{0}, y_{0})$ to the $\zeta-$axis is the point itself and we set $\zeta = 0$ at $(x_{0}, y_{0})$.}
\item{Step 5: To calculate the parameter $\zeta_{0}$ which gives $\phi (\zeta_{0}) = 0$
so that the hyperbolic tangent profile of $\phi$ along the $\zeta-$axis,
$\phi (\zeta) = - \tanh \frac{\zeta - \zeta_{0}}{W / 2}$, is fully determined.
Note that Newton's method is employed to solve the nonlinear equation to find $\zeta_{0}$.}
\item{Step 6: To calculate the order parameter $\phi$ at $(x_{vp}, y_{vp})$ as 
$\phi (x_{vp}, y_{vp}) = \phi (\zeta_{vp}) = - \tanh \frac{\zeta_{vp} - \zeta_{0}}{ W / 2} $.
The value of $\phi$ at $(x_{v}, y_{v})$ is obtained as $\phi (x_{v}, y_{v}) = \phi (x_{vp}, y_{vp})$
because the two points are on the same characteristic line.}
\end{itemize}
It is obvious that the above steps involve at most nine points (including $(x_{0}, y_{0})$ and its eight neighbours).
The neighbouring points in Group-S are actually just used to roughly determine the interface orientation.
In essence, the unique feature of the present method is due to Steps 4 \& 5.
By \emph{projecting the ghost point onto the $\zeta-$axis and employing the hyperbolic tangent profile of $\phi$ in this direction},
we circumvent the usual interpolation step to find $\phi (x_{v}, y_{v})$ in previous methods which may become quite complex 
as it involves the determination of the local interface configuration in the mesh lines~\cite{jcp15diim-curved}.
One might worry that the solution of the nonlinear equation brings additional cost.
According to our experience, the convergence in the solution process to find $\zeta_{0}$ by Newton's method is quite fast.
What is more, the above WBC is applied only in the interfacial region 
(set according to the criterion $\vert \phi (x_{0}, y_{0}) \vert < 0.99$)
which just covers a few grid points.
When $\phi (x_{0}, y_{0}) > 0.99$, the value of $\phi$ at the ghost point $\phi (x_{v}, y_{v})$ is simply set to be $1.0$,
and when $\phi (x_{0}, y_{0}) < -0.99$, it is set to be $-1.0$.
Therefore, the additional computational cost is quite small.
We would like to note that in actual implementation we did not use the element of the array $\phi$ at $(x_{v}, y_{v})$
to store the value obtained in the above way because at this position the point itself can also be a fluid node near the wall
(e.g,, on the other side of the wall when the wall is infinitely thin).
Our solution to this problem is to introduce a local array for the point at $(x_{0}, y_{0})$ to store the values
of $\phi$ at its eight neighbouring points (with some of them being the ghost points).
These values stored in the local array are used to evaluate the derivatives at $(x_{0}, y_{0})$.

\begin{figure}[htp]
  \centering
 \includegraphics[trim = 0mm 0mm 0mm 0mm, clip, scale = 1.0, angle = 0]{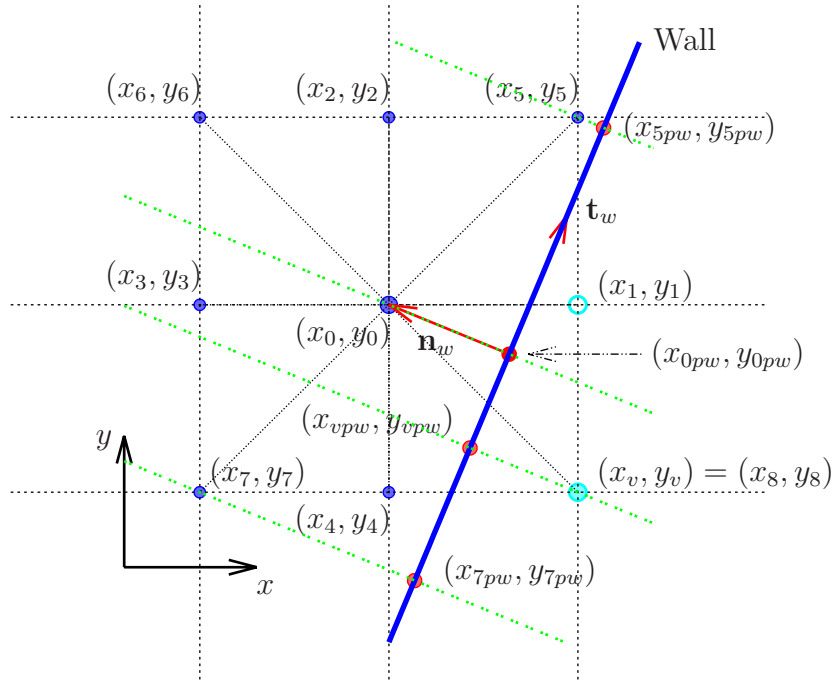}
  \caption{Illustration of the proposed implementation of the boundary condition for the chemical potential $\mu$
  on an immersed wall.}
  \label{fig:projection-wall}
\end{figure}

Figure \ref{fig:projection-wall} illustrates the implementation of the no-flux condition on a surface
($\frac{\partial \mu}{\partial n_{S}} \vert _{S}  = 0 $) for the chemical potential $\mu$.
There are some differences between this implementation for $\mu$ and the above for $\phi$
since the chemical potential does not follow the hyperbolic tangent profile 
and the condition for $\mu$ is independent of the contact angle $\theta_{w}$.
The implementation for $\mu$ involves more points surrounding $(x_{0}, y_{0})$ 
(but still limited to the nearest and next-to-nearest neighbours).
For convenience, the eight neighbours are labelled with the index $l$ ($l = 1, 2, \cdots, 8$)
corresponding to the D2Q9 lattice velocity model in LBM.
The specific steps to obtain $\mu (x_{v}, y_{v})$ are as follows.
\begin{itemize}
\item{Step 1: To project all nine points $(x_{l}, y_{l})$ ($l = 0, 1, \cdots, 8$) onto the wall to obtain nine points on the wall
$(x_{lpw}, y_{lpw})$ ($l = 0, 1, \cdots, 8$).}
\item{Step 2: Among the neighbouring points in Group-S 
search for the two indices with the maximum and minimum $x-$coordinates
upon the above projection onto the wall $x_{pw}$, denoted as $l_{\max}$ and $l_{\min}$
(for example, $l_{\max} = 5$ and $l_{\min} = 7$ in fig. \ref{fig:projection-wall}).
Note that without loss of generality we assume here the wall is \emph{not} vertical
(if it is vertical, the maximum and minimum $y-$coordinates $y_{pw}$ are considered instead).}
\item{Step 3: To calculate the coordinates on the $\eta-$axis (along the wall) for 
the three points $(x_{vpw}, y_{vpw})$, $(x_{l_{\max}pw}, y_{l_{\max}pw})$ and $(x_{l_{\min}pw}, y_{l_{\min}pw})$
to obtain $\eta_{v}$, $\eta_{l_{\max}}$ and $\eta_{l_{\min}}$ (note $\eta_{0} = 0$ is at $(x_{0pw}, y_{0pw})$).}
\item{Step 4: Upon assuming a quadratic polynomial for the variation of $\mu$ along the $\eta-$axis,
$\mu = a \eta^{2} + b \eta + c$,
to solve for the coefficients $a$, $b$, and $c$ from the three known conditions
$\mu (0) = \mu (x_{0pw}, y_{0pw}) \approx \mu (x_{0}, y_{0})$, 
$\mu (\eta_{l_{\max}}) = \mu (x_{l_{\max}pw}, y_{l_{\max}pw}) \approx \mu (x_{l_{\max}}, y_{l_{\max}})$,
and $\mu (\eta_{l_{\min}}) = \mu (x_{l_{\min}pw}, y_{l_{\min}pw}) \approx \mu (x_{l_{\min}}, y_{l_{\min}})$
where the condition $\frac{\partial \mu}{\partial n_{w}}  = 0$ has been used.}
\item{Step 5: The value of $\mu$ at $(x_{vpw}, y_{vpw})$ 
is found by substituting the coordinate $\eta_{v}$ into the polynomial, $\mu (\eta_{v}) = a \eta_{v}^{2} + b \eta_{v} + c$.
The value of $\mu$ at $(x_{v}, y_{v})$ is also obtained as 
$\mu (x_{v}, y_{v}) \approx \mu (x_{vpw}, y_{vpw})$.}
\end{itemize}

\section{Results and Discussions}\label{sec:res-dis}

In this section we present the study of two problems to verify the proposed method
to implement the wetting boundary condition.
The first is on the shape of a drop on a circular cylinder with different contact angles.
The second is about the spreading of a drop on a flat wall embedded in the domain. 
In the first problem, 
the main focus is on the final interface shape whereas in the second the dynamic evolution of the interface
will be examined.
Since our main purpose is to verify the above approach to implement the WBC,
we will fix most of the physical parameters and select some appropriate computational parameters.
Both the density ratio and the dynamic viscosity ratio are fixed to be unity.
The initial drop radius $R$ is chosen to be the reference length, $L_{r} = R$.
The reference velocity is derived from the interfacial tension $\sigma$ and dynamic viscosity $\eta$ 
as $U_{r} = \sigma / \eta$.
From $L_{r}$ and $U_{r}$, a reference time is derived as $T_{r} = L_{r} / U_{r} = L_{r} \eta / \sigma$.
All length and time quantities are measured in $L_{r}$ and $T_{r}$ respectively.
Based on these quantities, 
the capillary number can be calculated as $Ca = \eta U_{r} / \sigma = 1$
and the Reynolds number, 
$Re = \rho U_{r} L_{r} / \eta = \rho \sigma L_{r} / \eta^{2}$.
We fix the Reynolds number at $Re=100$.
The reference length $L_{r}$ 
is discretized into $N_{L}$ uniform segments
and the reference time $T_{r}$ 
is discretized into $N_{t}$ uniform intervals.
Thus, the grid size and time step are $\delta_{x} = L_{r} / N_{L}$ and $ \delta_{t} = T_{r} / N_{t}$.
The interface width measured in the grid size $W/\delta_{x}$ is fixed to be $4.0$.
In diffuse-interface simulations of binary fluids, there are two additional (numerical) parameters:
the Cahn number $Cn = W / L_{r}$
(i.e., the ratio of interface width over the reference length)
and the Peclet number $Pe = (U_{c} L_{c}^{2}) / (M \sigma) $
(i.e., the ratio of convection over diffusion in the CHE).
In all simulations we use $Pe=5000$.
The values of $N_{L}$, $N_{t}$, and $Cn$ will be given later for each problem.

\subsection{Drop on a circular cylinder}

\begin{figure}[htp]
  \centering
 \includegraphics[trim= 20mm 20mm 5mm 20mm, clip, scale = 1.3, angle = 0]{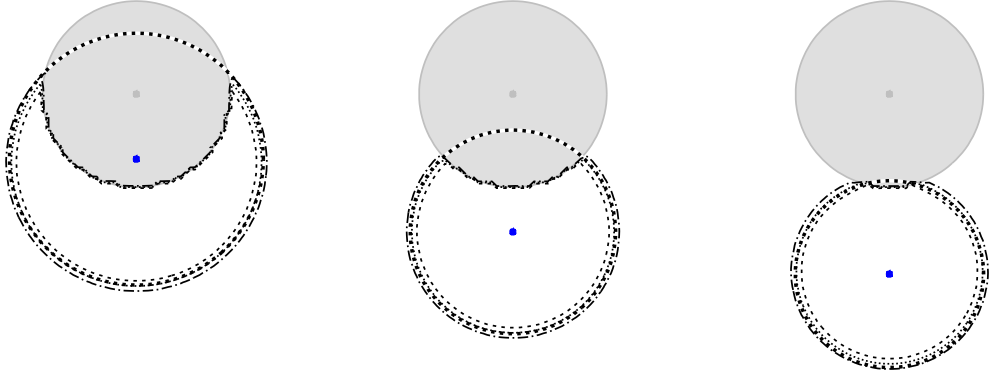}
  \caption{
  Equilibrium shapes of a drop on a circular cylinder with the contact angle of the outer surface being
  $30^{\circ}$, $90^{\circ}$, and $150^{\circ}$ (from left to right).
  The shaded circle denotes the circular cylinder.
  The three parallel lines are the contour lines for $\phi = -0.5$, $0$, and $0.5$.
  The dashed circle partly penetrating into the cylinder represents the theoretical prediction
  (the small blue dot denotes its center).
Some shared parameters for the simulations are $Cn=0.2$, $N_{L} = 20$, $N_{t} = 200$, $Re=100$.
}
    \label{fig:drop-circ-cylinder}
\end{figure}

\begin{figure}[htp]
  \centering
  \includegraphics[trim= 10mm 20mm 10mm 20mm, clip, scale = 1.25, angle = 0]{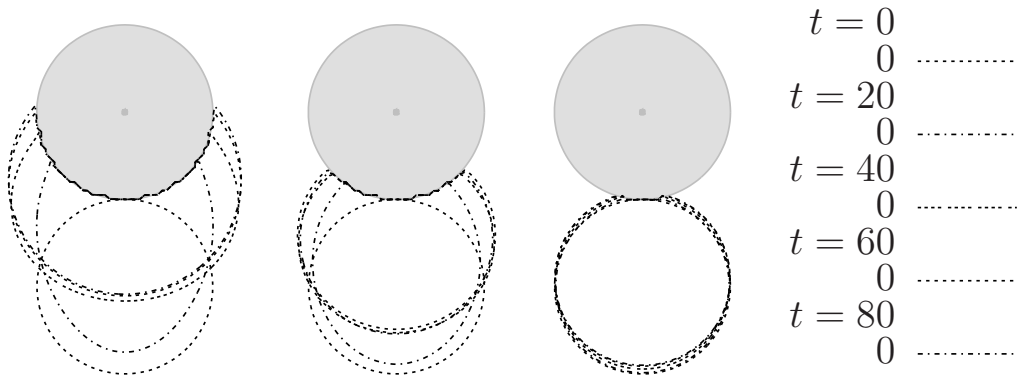}
  \caption{
  Snapshots of the interface shapes for a drop on a circular cylinder at different times before reaching equilibrium.
  The contact angles of the outer surface of the cylinder are $30^{\circ}$, $90^{\circ}$, and $150^{\circ}$ (from left to right).
  }
    \label{fig:drop-circ-cylinder-evol}
\end{figure}

\begin{figure}[htp]
  \centering
  \includegraphics[trim= 2mm 10mm 10mm 25mm, clip, scale = 1.25, angle = 0]{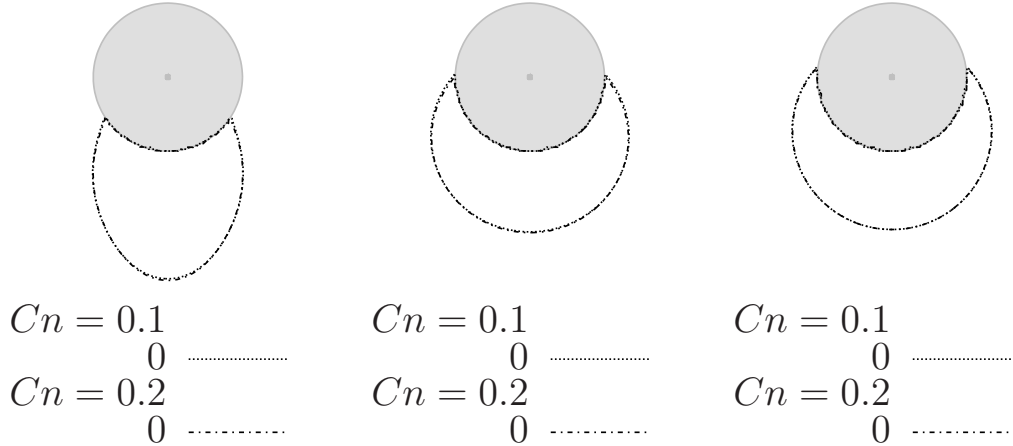}
  \caption{
  Snapshots of the interface shapes for a drop on a circular cylinder with a contact angle $30^{\circ}$ at different times 
  (from left to right: $t=20$, $60$, and $100$)
  simulated with two different resolutions: $Cn=0.2$, $N_{L} = 20$, $N_{t} = 200$
  and $Cn=0.1$, $N_{L} = 40$, $N_{t} = 400$.
  }
    \label{fig:ca030-drop-circ-cylinder-evol-convg}
\end{figure}

First, the results for a drop on a circular cylinder are presented.
In this problem, a circular cylinder having a radius $R_{cyl} = 1$ is fixed at $(x_{cyl}, y_{cyl}) = (2.0, 3.5)$.
The wetting property of the outer surface of the cylinder is characterized by a contact angle $\theta_{w}$.
Initially, the drop is below the cylinder and its center is located at $(x_{cd}, y_{cd}) = (2.0, 1.5)$,
which means that the drop contacts the cylinder by only one point.
The size of the domain is $4 \times 6$.
Other computational parameters are $N_{L}  = 20$, $N_{t} = 200$, and $Cn = 0.2$
for most of the following results (unless specified otherwise).
We tested three contact angles, $\theta_{w} = 30^{\circ}$, $90^{\circ}$, and $150^{\circ}$.
Figure \ref{fig:drop-circ-cylinder} shows the shapes of the drop at the end of the simulations ($t=100$)
when the interface nearly remains stationary.
Also shown in fig. \ref{fig:drop-circ-cylinder} are the shapes in equilibrium calculated by theory
(details of the theoretical prediction can be found in~\cite{jcp15diim-curved}).
It is seen that the final shapes agree with the theoretical predictions well for all three contact angles.
These results suggest that using the present method one can calculate the equilibrium interface shape
on a curved surface accurately.
Besides, in figure \ref{fig:drop-circ-cylinder-evol}
we plot the evolutions of the interface at an interval $\Delta t = 20$ before the equilibrium state is achieved. 
It is observed that the change of the interface every $\Delta t$ becomes smaller and smaller as time increases.
Finally, the convergence of the simulation is also briefly examined for $\theta_{w} = 30^{\circ}$
under which the interface motion is the strongest.
Figure \ref{fig:ca030-drop-circ-cylinder-evol-convg} compares the interface shapes for this case
at $t=20$, $60$, and $100$ simulated at two different resolutions:
one is as the above and the other uses finer spatial and temporal resolutions 
with $Cn=0.1$, $N_{L} = 40$, $N_{t} = 400$.
One can see from fig. \ref{fig:ca030-drop-circ-cylinder-evol-convg}
that the difference in the interfaces between the two simulations are quite small.
This seems to suggest that using the present method
the numerical solution for the interface movement on a curved surface
converges to some limit when $Cn$ decreases with $Pe$ being fixed
(note $N_{L}$ and $N_{t}$ should be changed properly).
This agrees with some of the previous findings on contact line motions on a flat surface 
by phase-field simulations~\cite{jfm10-sil-che-cl}.

\subsection{Drop spreading on an inclined wall}

\begin{figure}[htp]
  \centering
   \includegraphics[trim= 20mm 10mm 20mm 10mm, clip, scale = 0.8, angle = 0]{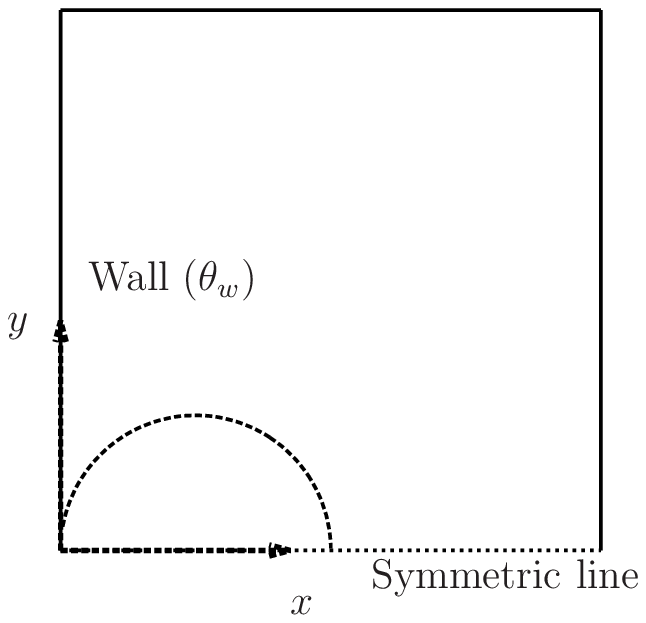}
    \includegraphics[trim= 10mm 10mm 10mm 10mm, clip, scale = 0.8, angle = 0]{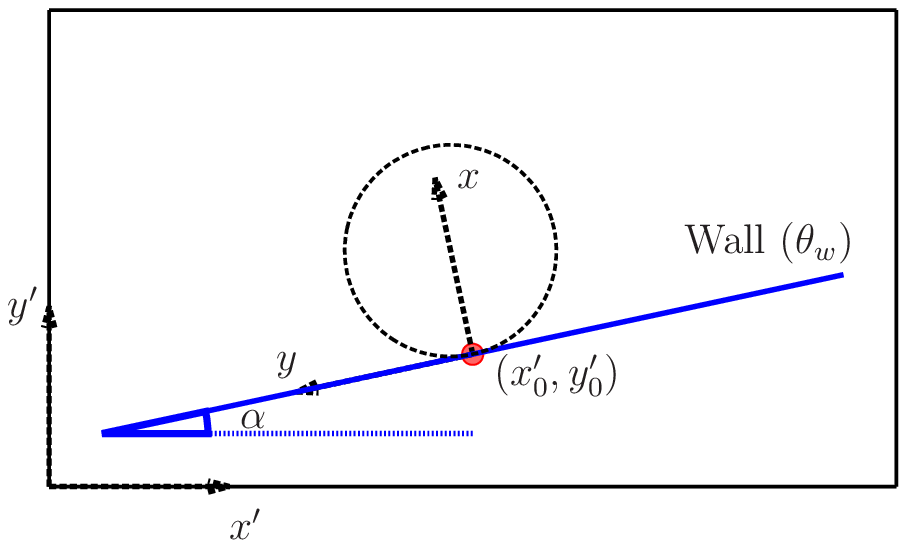}
  \caption{
  Initial setups for the drop spreading problem by two types of simulations.
  Left: a drop spreading on the left wall where the original characteristic interpolation is applied; 
  right: a drop spreading on the wall embedded inside the domain with an inclination angle $\alpha$ 
  on which the proposed method is applied.
  The dashed line represents the initial interface (the contour line for $\phi = 0$).
  The solid blue line represents the wall.
}
    \label{fig:drop-spreading-init-setup}
\end{figure}

\begin{figure}[htp]
  \centering
   \includegraphics[trim= 40mm 10mm 40mm 10mm, clip, scale = 0.8, angle = 0]{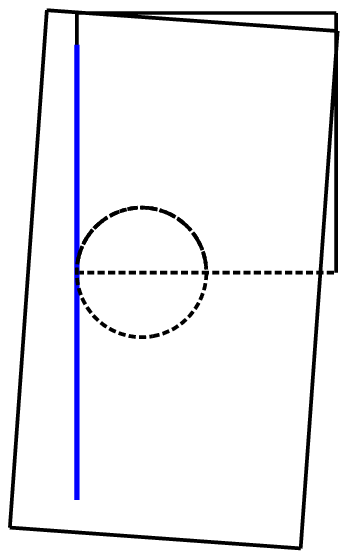}
  \includegraphics[trim= 40mm 10mm 40mm 10mm, clip, scale = 0.8, angle = 0]{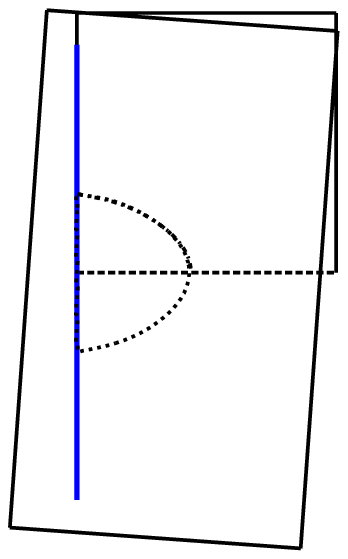}
   \includegraphics[trim= 40mm 10mm 40mm 10mm, clip, scale = 0.8, angle = 0]{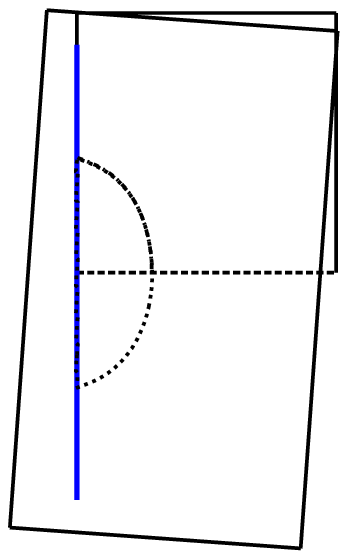}
    \includegraphics[trim= 40mm 10mm 40mm 10mm, clip, scale = 0.8, angle = 0]{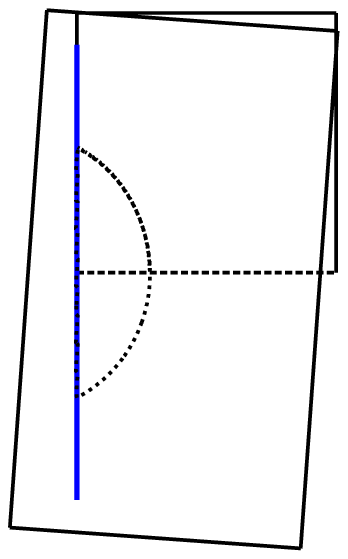}
 \includegraphics[trim= 40mm 10mm 40mm 10mm, clip, scale = 0.8, angle = 0]{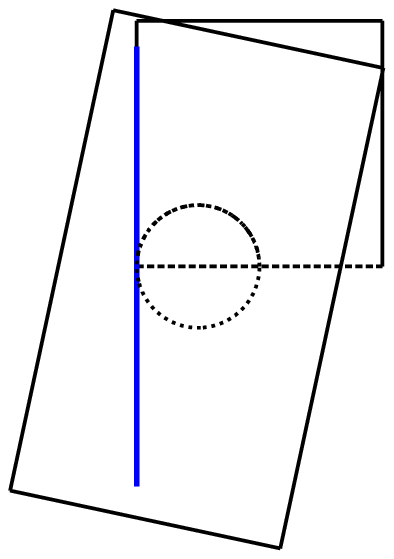}
  \includegraphics[trim= 40mm 10mm 40mm 10mm, clip, scale = 0.8, angle = 0]{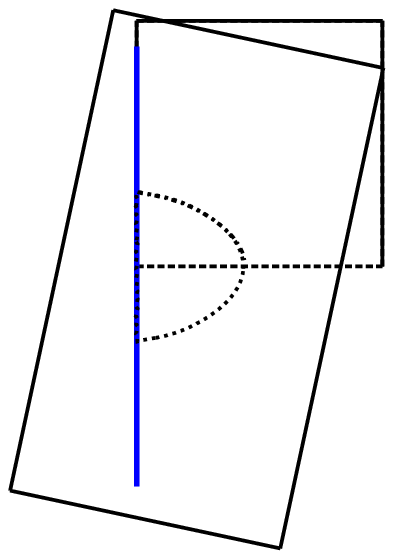}
   \includegraphics[trim= 40mm 10mm 40mm 10mm, clip, scale = 0.8, angle = 0]{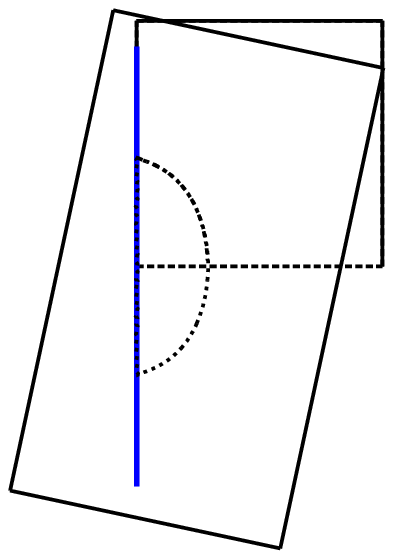}
    \includegraphics[trim= 40mm 10mm 40mm 10mm, clip, scale = 0.8, angle = 0]{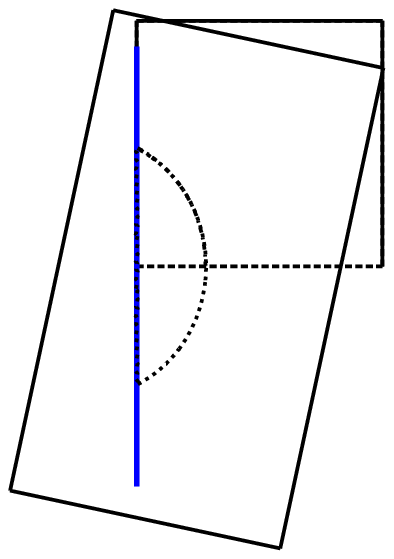}
  \caption{
  Snapshots of the interfaces at $t=0$, $20$, $40$, and $60$ (from left to right)
  by two types of simulations with one overlaid on the other.
  For the simulations using the proposed method, 
  the angle of inclination is $\alpha=4.086^{\circ}$ in the upper row and $\alpha=12.095^{\circ}$ in the lower row.
  Some shared parameters for the simulations are $Cn=0.125$, $N_{L} = 32$, $N_{t} = 128$, $Re=100$.
}
    \label{fig:drop-spreading}
\end{figure}

Next, we investigate the spreading of a drop on a wall placed inside the fluid domain at a given angle of inclination $\alpha$.
The wall is represented by a straight line segment between two points at $(x_{s,l}, y_{s,l})$ and $(x_{e,l}, y_{e,l})$.
Two values of $\alpha$ were tried: $\alpha=4.086^{\circ}$ and $12.095^{\circ}$ ($\tan \alpha = 1 /14$ and $3/14$).
The start point was fixed at $(x_{s,l}, y_{s,l}) = (0.5, 0.5)$ whereas
$(x_{e,l}, y_{e,l}) = (7.5, 1.0)$ was used for $\alpha=4.086^{\circ}$, 
and for $\alpha = 12.095^{\circ}$, $(x_{e,l}, y_{e,l}) = (7.5, 2.0)$.
The contact angle on this immersed wall is $\theta_{w} = 60^{\circ}$.
The domain size is $8 \times 4.5$.
Periodic boundary conditions are applied at all four domain boundaries.
It is noted that gravity effects are not included in this problem and the spreading is purely driven by interfacial tension forces.
For comparison, we also carried out another simulation using the characteristic interpolation (CI) given in~\cite{cf11-accur-cabc}.
In the simulation using the original CI, 
the problem is symmetric about the $x-$axis and only the upper half of the domain is used 
(the domain size for simulation is $4 \times 4$);
except for the bottom boundary, the boundary conditions for a static wall are applied at the other three boundaries.
Figure \ref{fig:drop-spreading-init-setup} illustrates the initial setups of the two different kinds of simulations.
The left figure is for the simulation using the original CI, and the right is for the current simulation.
The small red dot at $(x^{\prime}_{0}, y^{\prime}_{0})$ in the right figure
denotes the initial point of contact between the drop and the wall.
In the simulation at $\alpha=4.086^{\circ}$, 
the initial drop center $(x_{cd}, y_{cd})$ is located at $(3.9288, 1.7475)$, 
and the contact point is at $(x^{\prime}_{0}, y^{\prime}_{0}) = (4, 0.75)$.
In the simulation at $\alpha=12.095^{\circ}$, $(x_{cd}, y_{cd})=(3.7905, 2.2278)$,
and $(x^{\prime}_{0}, y^{\prime}_{0}) = (4, 1.25)$.
It is also noted that the drop spreading studied here is relatively slow 
and the boundaries of the domain have little effect on the spreading process.
Some shared computational parameters are $N_{L}  = 32$, $N_{t} = 128$, and $Cn = 0.125$.
Figure \ref{fig:drop-spreading} compares the evolutions of the interface by the simulation using CI
and the current simulations at the two inclination angles.
Note that in order to compare the interface shapes, 
the coordinates in the current simulations have been transformed as: 
$x = (x^{\prime} - x^{\prime}_{0}) \cos \beta + (y^{\prime} - y^{\prime}_{0}) \sin \beta$,
$y = - (x^{\prime} - x^{\prime}_{0}) \sin \beta + (y^{\prime} - y^{\prime}_{0}) \cos \beta$,
where $\beta = \alpha + \frac{\pi}{2}$.
It is found from fig. \ref{fig:drop-spreading} that at both values of $\alpha$
the differences between the present interface shapes and those using CI are quite small at all times.
This means that the proposed method can be used to simulate dynamic problems involving contact line movement
with the outcomes matching the corresponding simulations using the original CI.


\section{Concluding Remarks}\label{sec:conclusion}

To conclude, we have proposed and verified a new geometric formulation for the wetting boundary condition 
on general surfaces immersed inside the domain
for two-phase flow simulations using the Cahn-Hilliard equation.
This formulation is simple in the sense that only one point in the fluid is directly required to find 
the unknown value of the order parameter at a ghost point.
At the same time, the accuracy is well maintained with the computational cost being only marginally higher.
The special feature makes it quite attractive especially when curved immersed surfaces are encountered
since there is no need to determine the local configuration of the interface with respect to the grid. 
Another minor advantage is that in parallel computation based on domain decomposition using MPI
only one layer of information need to be exchanged between different computational nodes
because only the nearest and next-to-nearest neighbouring points are required in the present method.
Future work includes its extension to three dimensions.

\bf Acknowledgement \rm

This work is supported by the National Natural Science Foundation of China 
(NSFC, Grant No. 11202250).

\bibliography{/Users/jjhuang/work/bib/MyReference}

\end{document}